\documentclass[twocolumn,nofootinbib,amsmath,amssymb,a4paper]{revtex4}

\usepackage{graphicx}
\usepackage{dcolumn}
\usepackage{bm}

\usepackage{relsize}

\def\mbc{$M^{}_{\rm bc}$}
\def\mmbc{M^{}_{\rm bc}}
\def\deltaE{$\Delta E$}

\def\phitwo{$\phi^{}_2$}

\def\ra{\!\rightarrow\!}
\def\bbar{\overline{B}{}^{\,0}}

\def\brhorho{$B^0\ra\rho^+\rho^-$}

\def\babar{\mbox{\slshape B\kern-0.1em{\smaller A}\kern-0.1em
    B\kern-0.1em{\smaller A\kern-0.2em R}}}

\begin{document}


\title{Measurement of the CKM angle $\alpha$ ($\phi_2$) with 
$B\rightarrow \rho\rho$  decays at Belle and \babar~\footnote{Talk
presented at CKM2006, Nagoya, Japan, December 12-16 2006. University of Cincinnati preprint UCHEP-07-03.}} 

\author{A. Somov}
 \email{somov@physics.uc.edu}
\affiliation{%
University of Cincinnati, Cincinnati, Ohio, 45221, USA\\
}%

\begin{abstract}
We overview recent measurements in $B\rightarrow \rho\rho$ decays
which are based on data samples collected at the PEP-II and KEKB  
asymmetric-energy $e^+e^-$ colliders with the \babar\ and Belle detectors.
Special emphasis is given to the determination of the $CP$-violating 
coefficients ${\cal A}$ and  ${\cal S}$ from an analysis of 
$B^0\rightarrow \rho^+\rho^-$ decays. The values of ${\cal A}$ and  ${\cal S}$,
branching fractions, and longitudinal polarization fractions of 
$B\rightarrow \rho\rho$ decays are used to constrain the Cabibbo-Kobayashi-Maskawa 
phase $\alpha (\phi_2)$ using an isospin analysis; the solution consistent with
the standard model is $71^\circ\!<\!\alpha (\phi^{}_2)\!<\!113^\circ$ at 68\%~C.L. 
 
\end{abstract}

\maketitle

\section{Introduction}
$CP$ violation in the Standard Model can be explained by the presence of 
an irreducible complex phase in the Cabibbo-Kobayashi-Maskawa~\cite{ckm} 
(CKM) quark-mixing matrix. The unitarity of the CKM matrix leads to six
triangles in the complex plane. One such triangle is given by
the following relation among the  matrix elements: 
$V^*_{ub}V_{ud} + V^*_{cb}V_{cd} + V^*_{tb}V_{td} = 0$.
The phase angle  $\phi_2$\footnote{The naming convention:  angles $\alpha$, $\beta$, 
and $\gamma$ used in \babar\ are referred to as $\phi_2$, $\phi_1$, and $\phi_3$ 
in Belle.}, defined as $arg[-(V_{td}V_{tb}^*)/(V_{ud}V_{ub}^*)]$, 
can be determined by measuring a time-dependent $CP$ asymmetry in 
$b\rightarrow u\overline{u}d$ decays such as 
$B^0\ra\pi^+\pi^-,\,\rho^\pm\pi^\mp$, 
and $\rho^+\rho^-$. The time-dependent decay rate for $B\rightarrow 
\rho^+\rho^-$  decays tagged with $B^0$($q=1$) and $\bbar$ ($q = -1$) mesons is 
given by
\begin{eqnarray}
\mathcal{P}_{\rho\rho} (\Delta t) =  \frac{e^{-|\Delta 
t|/\tau_{B^0}}}{4\tau_{B^0}}
\{ 1  + q[{\cal A}_{\rho\rho}\cos(\Delta m \Delta t) \\
 + {\cal S}_{\rho\rho}\sin(\Delta m \Delta t)]\},\nonumber
\label{eqn:rate}
\end{eqnarray}
where $\tau_{B^0}$ is the $B^0$ lifetime, $\Delta m$ is the mass 
difference between 
the two $B^0$ mass eigenstates, $\Delta t = t_{CP} - t_{\rm tag}$, and 
${\cal A}_{\rho\rho}$ (\babar's definition is ${\cal C}_{\rho\rho} = -{\cal A}_{\rho\rho}$)
and ${\cal S}_{\rho\rho}$ are $CP$ asymmetry coefficients to be obtained 
from a fit to the experimental data. If the decay amplitude is a pure $CP$-even 
state  and is dominated by a tree diagram, 
 ${\cal S}_{\rho\rho} = {\rm sin}(2\phi_2)$ and ${\cal A}_{\rho\rho} = 0$. 
The presence of an amplitude with a different weak phase (such as from a "penguin" 
diagram) gives rise to direct $CP$ violation and shifts ${\cal S}_{\rho\rho}$ 
from ${\rm sin}(2\phi_2)$. However, the size of the loop amplitude is constrained 
by the branching fraction of $B^0\ra\rho^0\rho^0$~\cite{babar_rho0rho0}, indicating 
that this effect is small.

The $CP$-violating parameters receive contributions from a longitudinally 
polarized state ($CP$-even) and two transversely  polarized states (an admixture 
of $CP$-even and $CP$-odd states). Recent measurements of the polarization fraction by 
Belle and \babar~\cite{belle_rhorho,babar_alpha} show that the longitudinal polarization 
fraction is approximately $100\%$ ($f_L=0.968\pm 0.023$~\cite{hfag}). 
$f_L$ can be  extracted from a fit to helicity-angle distribution. The angular decay 
rate $d^2\Gamma / (\Gamma\,d\!\cos\theta_+\,d\!\cos\theta_-)$ is given by
$
\frac{9}{4} \left \{f_L \cos^2 \theta_+ \cos^2 \theta_- + \frac{1}{4}
(1-f_L)\sin^2\theta_+\sin^2\theta_- \right \},
$
where, $\theta^{}_{\!\pm}$ is the angle between the 
direction of the $\pi^0$ from the $\rho^\pm$ and the 
negative of the $B^0$ momentum in the $\rho^\pm$ rest frame.
\vspace{-0.7cm}

\section{Measurements}
The common features of $B\ra \rho\rho$ analyses are: (1) relatively small signal 
yields; the branching fractions of  $B\ra \rho\rho$ decays are in the order 
of $10^{-6}-10^{-5}$, (2) large width of $\rho$ mesons results in the large 
background; the fraction of signal events in most analysis is less than $1\%$, 
(3) there are several background sources: $e^+e^-\!\ra q\bar{q}\ (q=u,d,s,c)$ 
continuum, $b\!\ra c$, and $b\!\ra u$ backgrounds, (4) significant amount of 
events with multiple reconstructed $B$ candidates, (5) various variables are used 
in the likelihood  functions to distinguish among signal and backgrounds.

Both Belle and \babar\ analyses identify $B\rightarrow \rho\rho$ decays using the 
beam-energy constrained mass $\mmbc\!\equiv\!\sqrt{E^2_{\rm beam}-p^2_B}$ (called as 
beam-energy-substituted mass, $m_{ES}$, in \babar) and energy difference 
$\Delta E\!\equiv\!E^{}_B-E^{}_{\rm beam}$, where $E^{}_{\rm beam}$ is the beam 
energy, and $E^{}_B$ and $p^2_B$ are the energy and momentum of the reconstructed 
$B$  candidate, all evaluated in the center-of-mass (CM) frame.

The dominant background originates from  $e^+e^-\!\ra q\bar{q}\ (q=u,d,s,c)$
continuum events. To separate $q\bar{q}$ jet-like events from spherical-like
$B\overline{B}$ events, Belle uses event-shape variables, specifically, modified
Fox-Wolfram moments combined into a Fisher discriminant~\cite{KSFW}, and  
$\theta^{}_B$, the polar angle in the CM frame between the $B$ direction and the 
beam axis. The Fisher discriminant and $\theta^{}_B$ are used to form a ratio of 
signal and background likelihoods ${\cal R}$. In the $B^\pm \rightarrow \rho^\pm\rho^0$ 
analysis Belle also requires $|\cos\theta_T|<0.8$, where $\theta_T$ is the angle 
between the thrust axis of the candidate tracks and that of the remaining tracks 
in the event. In the \babar\ analyses $q\bar{q}$ background is suppressed by 
requiring $|\cos\theta_T|<0.8$ and making use of a neural network discriminant 
${\cal N}$ which is based on several event-shape variables. Bellow we describe 
$B\ra \rho\rho$ measurements in detail.
\begin{table}
\caption{\label{tab:rhopm} Reconstruction requirements used in 
$B^0\rightarrow \rho^+\rho^-$ analysis. The barrel and (endcap) regions in the 
electromagnetic calorimeter at Belle are defined as $32^\circ\!<\!\theta\!<\!129^\circ$ 
and ($17^\circ\!<\!\theta\!<\!32^\circ$ and $129^\circ\!<\!\theta\!<\!150^\circ$), 
where $\theta$ denotes the polar angle with respect to the beam axis.}
\renewcommand{\arraystretch}{1.1}
\begin{ruledtabular}
\begin{tabular}{lcc}
Cut & \babar\ & Belle \\
\hline
$E_\gamma$ (MeV)       &  50   &  50  (barrel)
 90 (endcap)      \\
$M^{}_{\gamma\gamma}$ (${\rm MeV}/c^2$) 
                       & 100. - 160.  & 117.8 - 150.2   \\
$ M_{\pi^\pm\pi^0}$   (${\rm GeV}/c^2$)
                       &  0.5 - 1.0   & 0.62 - 0.92     \\
$ \cos\theta^{}_{\!\pm}$ &   -0.9 - 0.98  & -0.8 - 0.98  \\  
$\mmbc$  (${\rm GeV}/c^2$) &  5.25 - 5.29  &  5.23 - 5.29  \\
                         &               &  5.27 - 5.29 ($CP$ fit) \\
$\Delta E$  (${\rm GeV}$)  &  -0.12 - 0.15 &  -0.2 - 0.3 \\
                         &               &  -0.12 - 0.08 ($CP$ fit) \\
\end{tabular}
\end{ruledtabular}
\vspace{-0.2cm}
\end{table}
\begin{figure}
\includegraphics[width=0.4\textwidth]{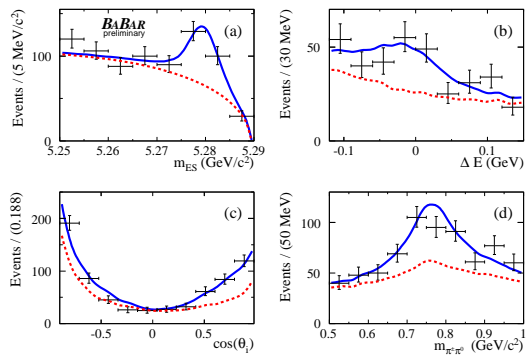}
\caption{\label{fig:babar_rhopm1} \babar\: (a) $m_{ES}$, (b) $\Delta E$, (c) $\cos\theta$, and 
(d) $m_{\pi^\pm\pi^0}$ for the highest purity tagged events. The curves show fit
projections: the dashed line is background and the solid line is the total.}
\vspace{-0.5cm}
\end{figure}
\begin{figure}
\includegraphics[width=0.3\textwidth]{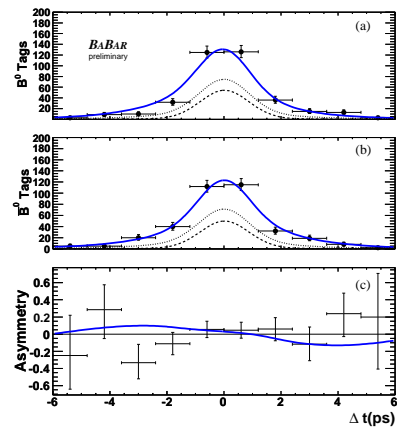}
\caption{\label{fig:babar_rhopm2} \babar\: the $\Delta t$ distributions of a signal-enriched 
sample for (a) $B^0$  and (b) $\bbar$ tagged events. (c) shows raw asymmetry. The dashed 
lines denote $q\bar q$ background, the dotted lines are the sum of background, and the solid 
lines are the total. }
\vspace{-0.3cm}
\end{figure}
\begin{figure}
\includegraphics[width=0.4\textwidth]{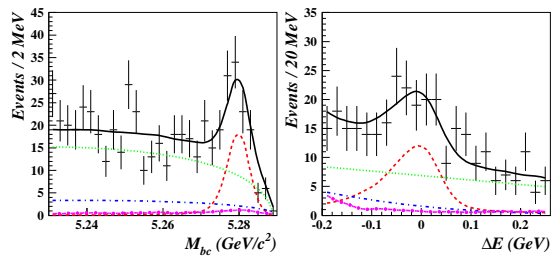}
\caption{\label{fig:belle_rhopm1} Belle: projections in $M_{\rm bc}$ (left) and $\Delta E$ 
(right) for the high-purity tagged events of a sample enriched in signal. The curves show 
fit projections: dashed is $\rho^+\rho^-\!+\rho\pi\pi$, dotted is $q\bar{q}$, dot-dashed 
is $b\ra c$, long-dashed is $b\ra u$, and solid is the total.}
\vspace{-0.5cm}
\end{figure}
\begin{figure}
\includegraphics[width=0.3\textwidth]{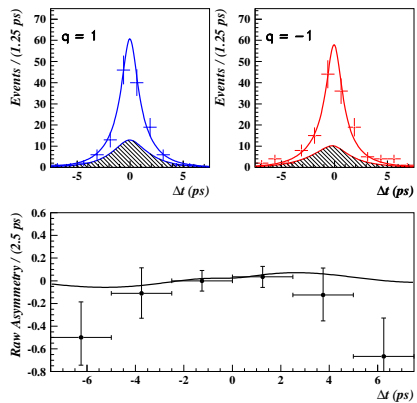}
\caption{\label{fig:belle_rhopm2} Belle: the $\Delta t$ distribution and projections of 
the fit for high-purity tagged events: (a) $B^0$ tags, (b) ${\overline B}^0$ tags. The raw 
$CP$ asymmetry is shown in (c). The hatched region shows signal events.}
\vspace{-0.5cm}
\end{figure}
\begin{figure}
\includegraphics[width=0.2\textwidth]{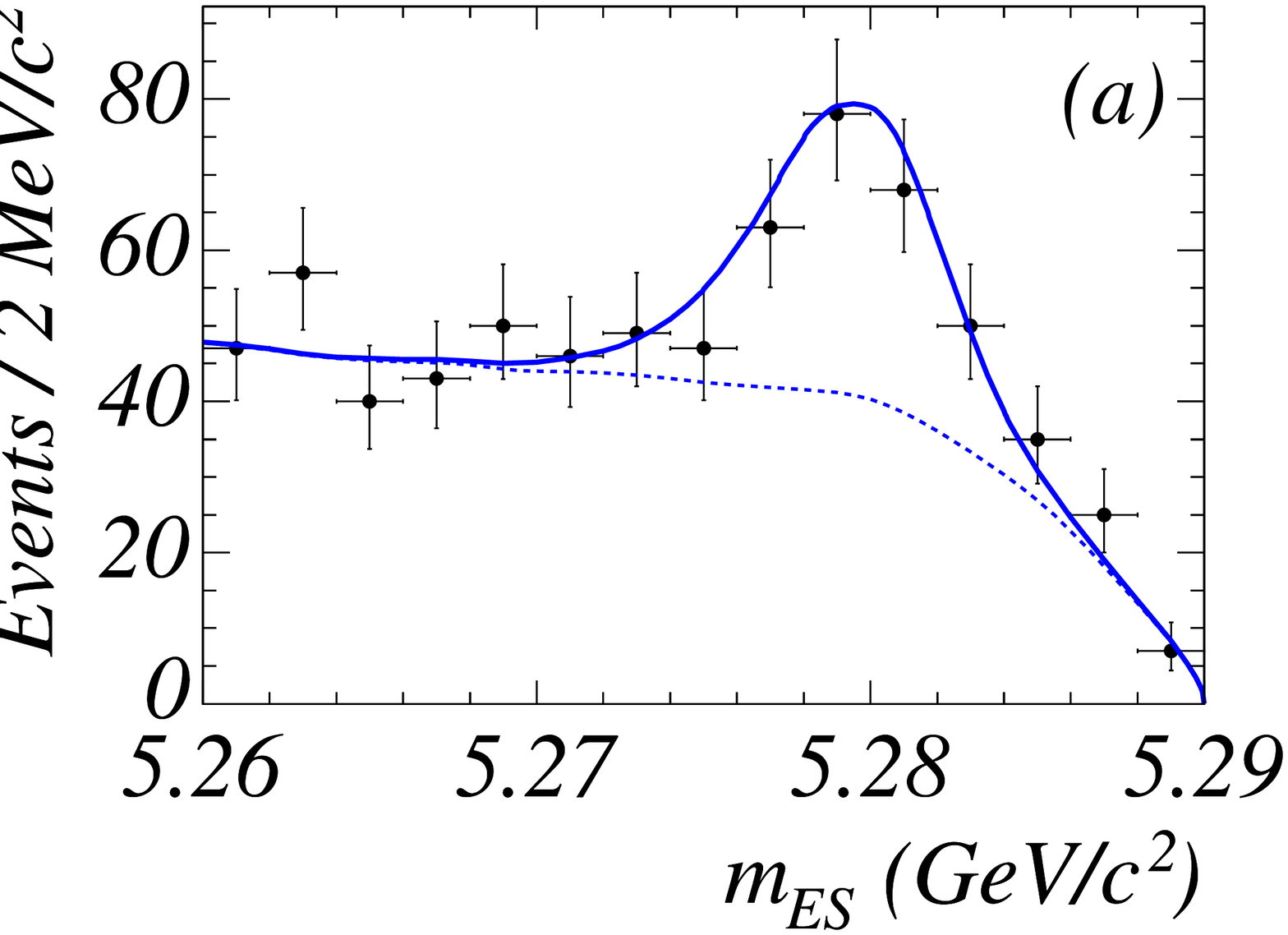}
\includegraphics[width=0.2\textwidth]{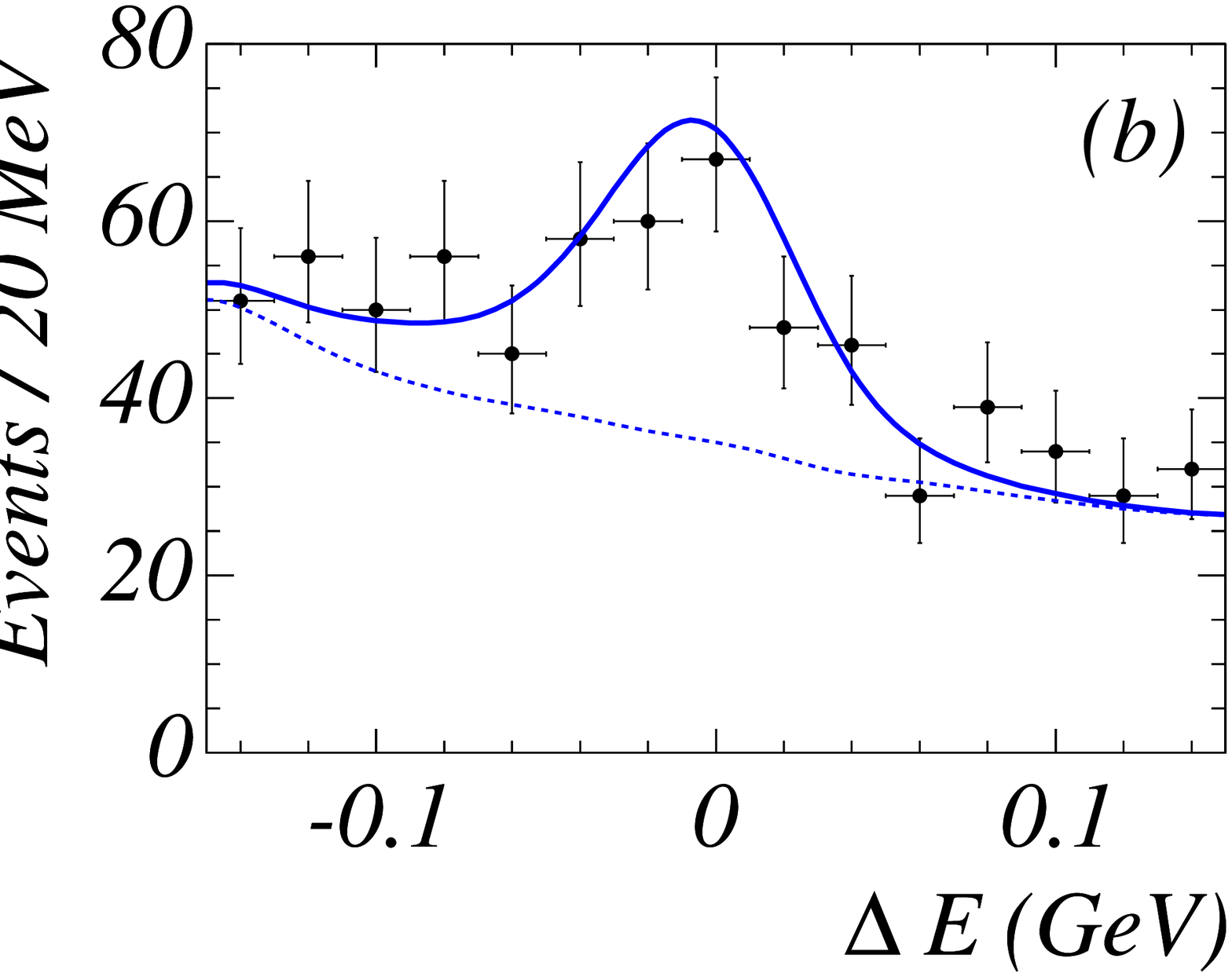}
\includegraphics[width=0.2\textwidth]{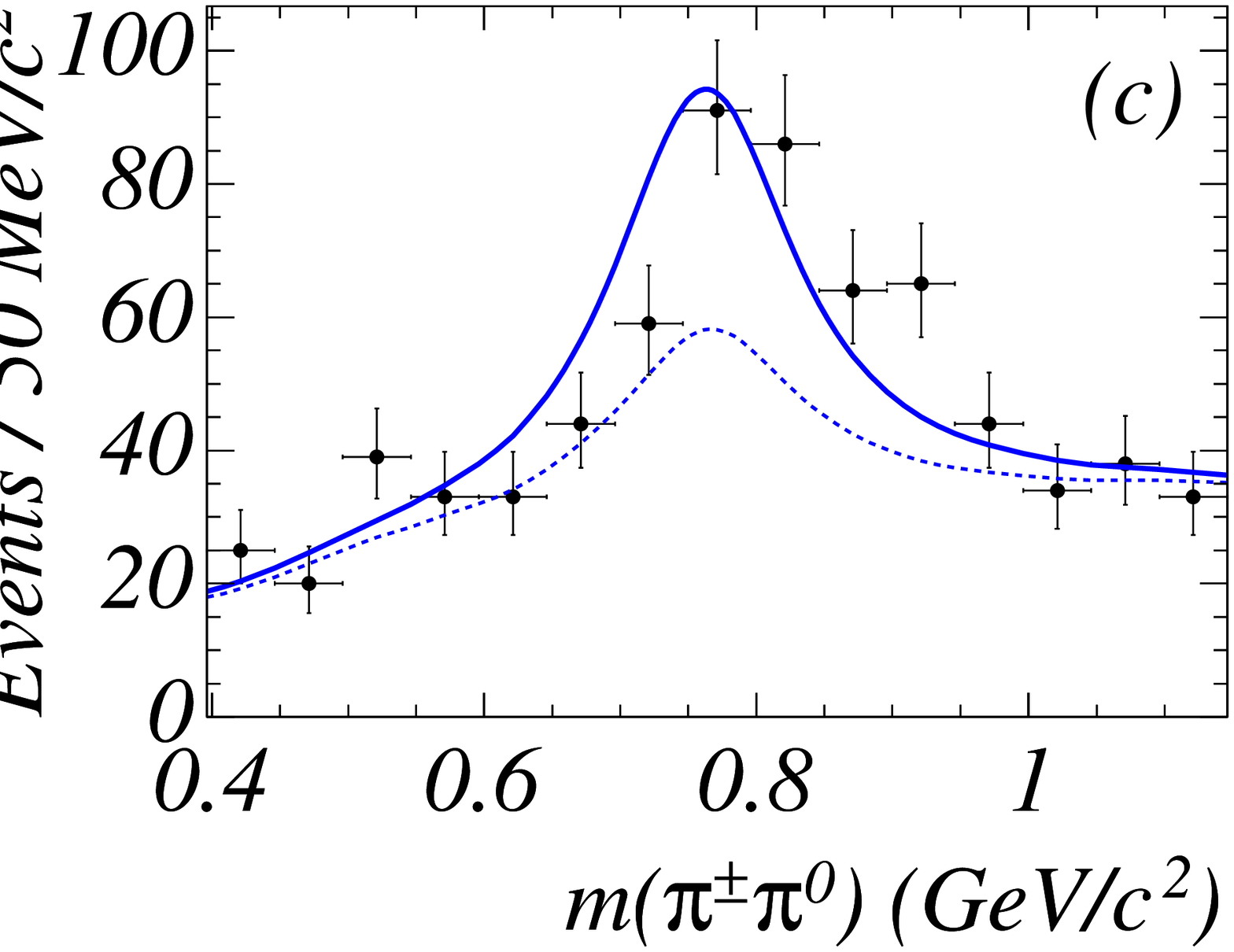}
\includegraphics[width=0.2\textwidth]{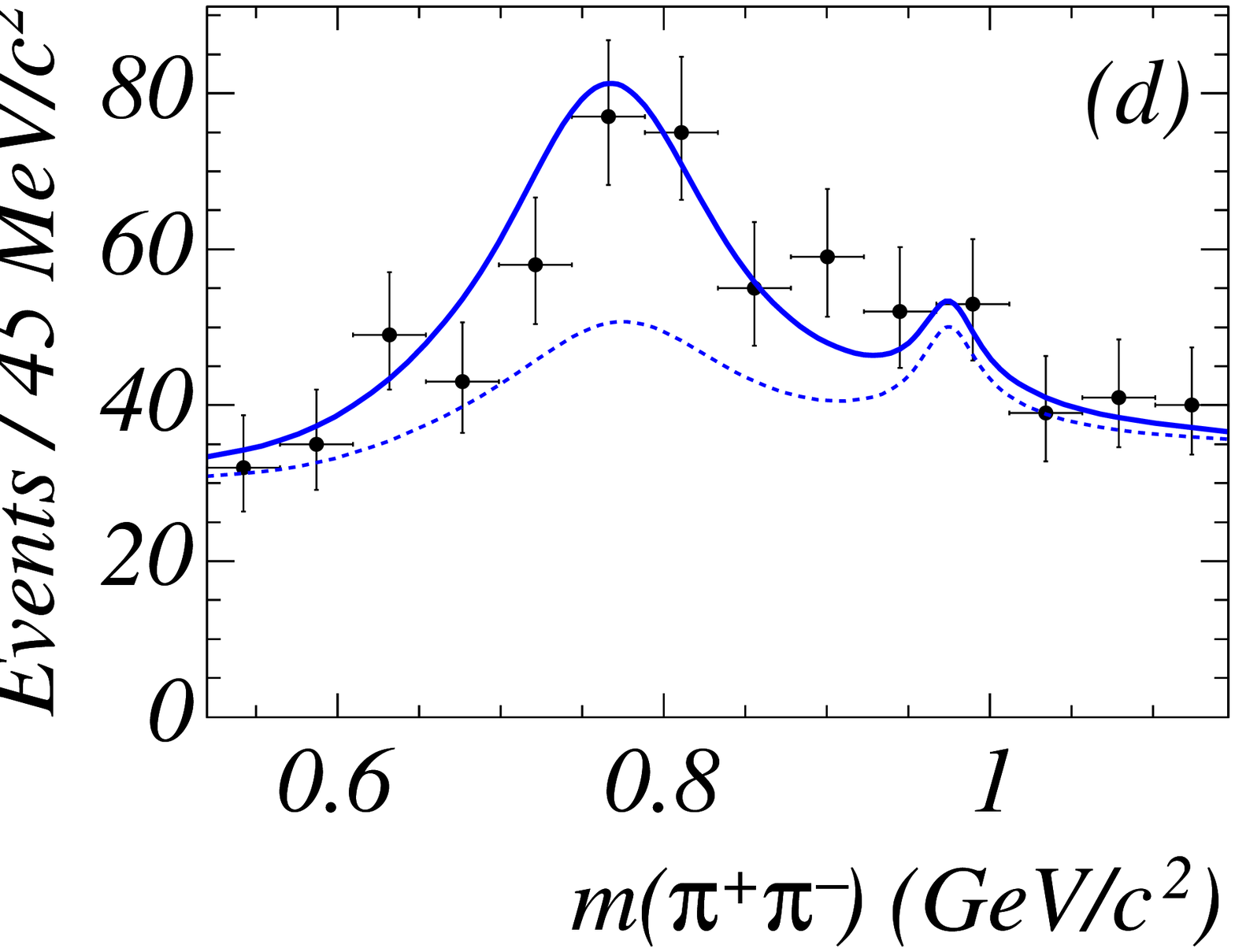}
\caption{\label{fig:babar_rhop} \babar\: (a) $m_{ES}$, 
(b) $\Delta E$, (c) $\cos\theta$, and (d) $m_{\pi^\pm\pi^0}$ distributions 
for a signal-enriched samples along with fit projections: the dashed lines 
show $q\bar q$ and $B\overline B$ backgrounds, the solid line is the total.}
\vspace{-0.5cm}
\end{figure}
\vspace{-0.2cm}

\subsection{$B^0 \rightarrow \rho^+\rho^-$}
\brhorho decays are reconstructed by combining two oppositely charged pion tracks with 
two neutral pions. The $\rho^\pm$ mesons are selected combining $\pi^\pm$ with $\pi^0$ 
candidates. The $\pi^0$ candidates are reconstructed from $\gamma\gamma$ pairs. Main 
event reconstruction requirements are listed in Table~\ref{tab:rhopm}. A flavor of the 
$B$ meson accompanying the \brhorho\ candidate is identified via its decay products. 
Tagging algorithms  yield the flavor of the tagged meson and a flavor-tagging quality. 
In events with multiple reconstructed $B$ candidates the best candidate is selected 
based on the $\pi^0$ masses, i.e. minimizing
$\sum_{\pi^0_{1,2}}(m_{\gamma\gamma} - m_{\pi^0})^2$. The fraction of signal decays 
which have at least one $\pi^\pm$ track incorrectly identified but pass all selection 
criteria is $13.8\%$ and $6.5\%$ for \babar\ and Belle, respectively. These are reffered
to as ``self-cross-feed'' (SCF) events. The following components are distinguished in the 
analyses: signal and $\rho\pi\pi$ non-resonant decays, SCF events,  continuum background 
($q\bar q$), charm $B$  background ($b\!\rightarrow\! c$), and charmless ($b\to u$)  background.
The $b\ra u$ background is dominated by 
$B\ra (\rho\,\pi,\,a^{}_1\pi,\,a^{}_1\,\rho,\rho^\pm\,\rho^0)$ decays.

\babar\ obtains the signal yield, $f_L$, and $CP$-violating parameters 
${\cal A}_{\rho^+\rho^-}$ and ${\cal S}_{\rho^+\rho^-}$ from an unbinned extended  
maximum-likelihood (ML) fit to $m_{ES}$, $\Delta E$, $\Delta t$, $m_{\pi^\pm\pi^0}$, 
$\cos\theta_\pm$, and ${\cal N}$ distribution of 33902 events~\cite{babar_rhopm_new}. 
\begin{figure}
\includegraphics[width=0.2\textwidth]{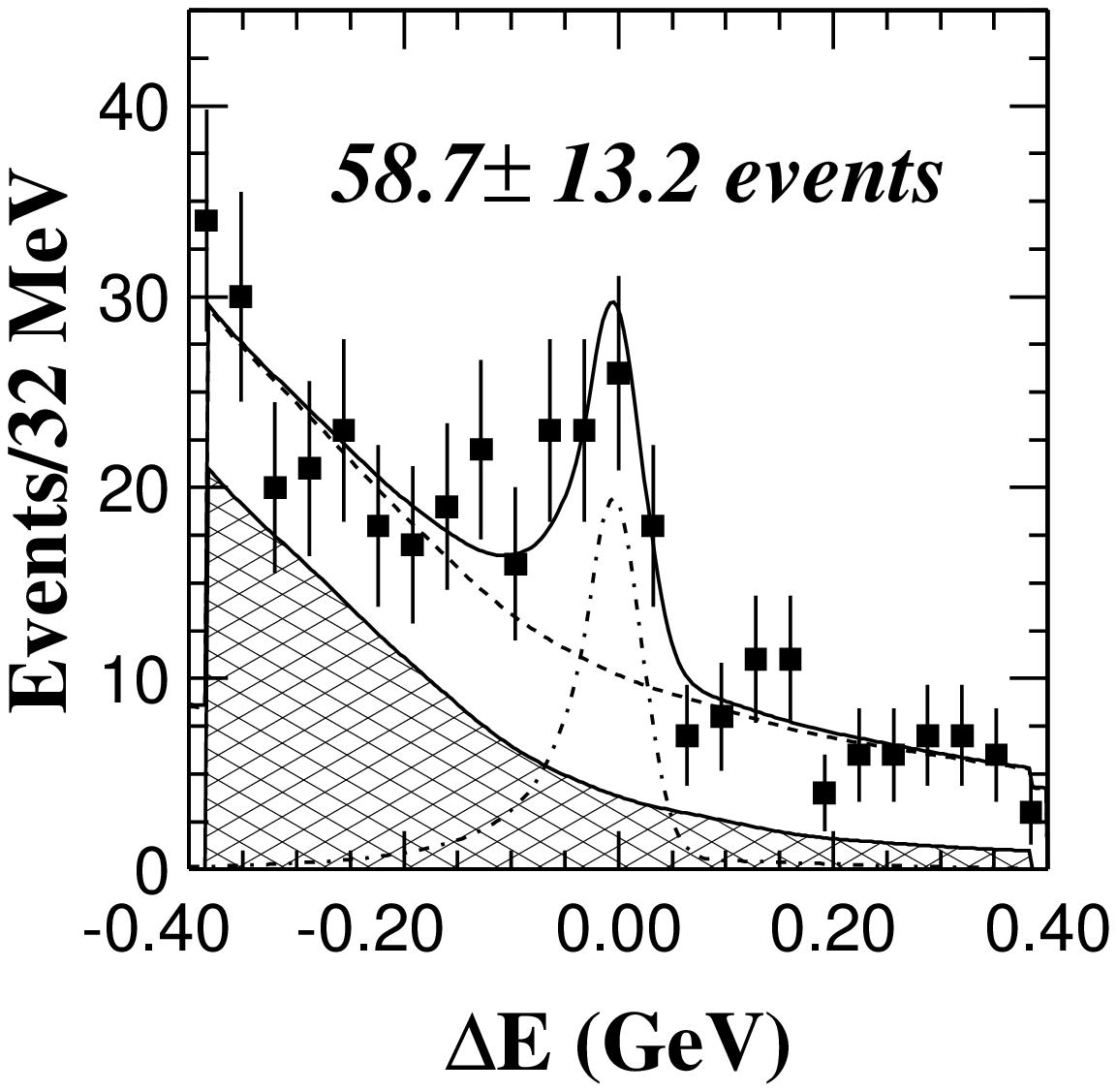}
\includegraphics[width=0.2\textwidth]{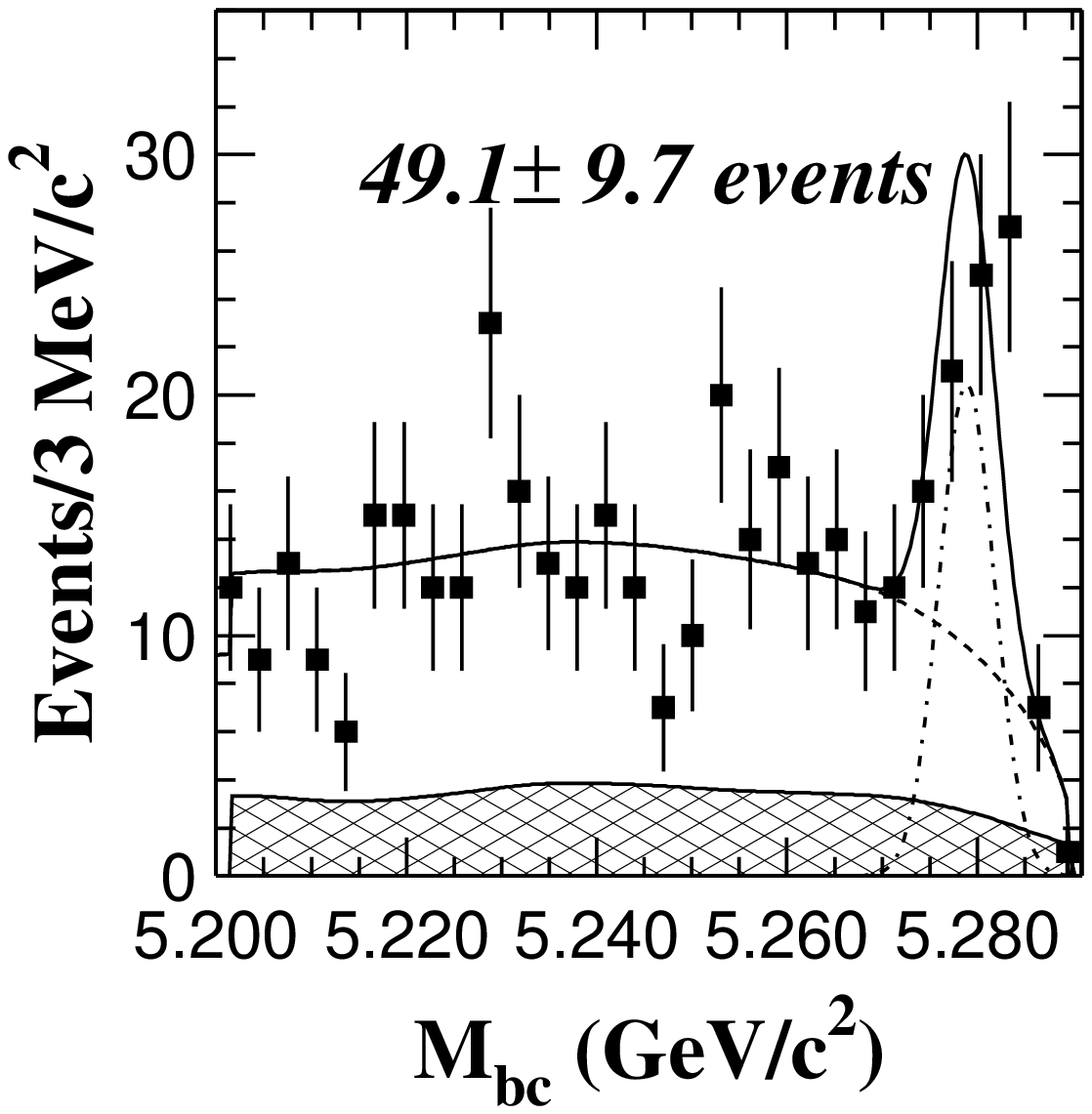}
\caption{\label{fig:belle_rhop} Belle: $\Delta E$ (left) and $M_{\rm bc}$ 
(right) distributions. The dashed curve is the sum of $B\overline B$ and 
continuum backgrounds, the dot-dashed curve is the signal, the hatched 
region shows $B\overline B$, and the solid curve is the total.}
\vspace{-0.6cm}
\end{figure}

The Belle analysis is organized in two main steps: (a) we first determine the
yields of signal and background components from an unbinned extended ML fit to 
the three-dimensional \mbc\ - \deltaE\  - ${\cal R}$ distribution. 
(b) we perform a fit to the $\Delta t$ distribution of 18004 events to determine the $CP$ 
parameters ${\cal A}_{\rho^+\rho^-}$ and  ${\cal S}_{\rho^+\rho^-}$
The fit results are presented in  Figures~\ref{fig:babar_rhopm1},~\ref{fig:babar_rhopm2},
~\ref{fig:belle_rhopm1}, and~\ref{fig:belle_rhopm2} and are listed in Table~\ref{tab:summary}.
The fraction of longitudinal polarization and fraction of non-resonant events ($6.3\pm6.7\%$) 
were measured in our previous analysis~\cite{belle_rhorho}. The branching fraction, $f_L$, and 
$CP$ asymmetries measured by Belle are similar to those obtained by \babar. The values 
${\cal A}_{\rho^+\rho^-}$ and ${\cal S}_{\rho^+\rho^-}$ are also consistent with no $CP$ 
violation (${\cal A} = {\cal S} = 0$).

\subsection{$B^\pm \rightarrow \rho^\pm\rho^0$}
The main reconstruction features of the analysis are the same as those for the
$B^0\rightarrow\rho^+\rho^-$.  Event selection requirements are listed in Table~\ref{tab:rhop}.

\babar\ obtained the yields of $B^\pm \rightarrow \rho^\pm\rho^0$ and $B^\pm \rightarrow \rho^\pm f_0$ 
decays, polarization, and charge asymmetry 
${\cal A}_{\rho^\pm\rho^0} = (N_{B^-} - N_{B^+})/(N_{B^-} + N_{B^+})$ using an unbinned extended 
ML fit to  $m_{ES}$, $\Delta E$, $\Delta t$, $m_{\pi^\pm\pi^0}$, 
$m_{\pi^+\pi^-}$, $\cos\theta_\pm$, $\cos\theta_0$, and ${\cal N}$ distribution of 74293 
events~\cite{babar_rhop0}. The charmless $b\rightarrow u$ background is dominated
by $\eta^\prime\rho^\pm$, $K^{*0}\rho^\pm$, $a_1^0\pi^\pm$, $a_1^\pm\pi^0$, $a_1^\pm \rho^0$, 
and $a_1^0\rho^\pm$ decays. 

The signal yield in Belle analysis~\cite{belle_rhop0} is obtained from a fit to $\Delta E$ 
distribution. To measure the polarization, Belle bins in $\cos\theta_\pm$ and $\cos\theta_0$ 
and determine the signal yield for each bin from the fit to $\Delta E$ distribution.
The polarization is obtained from a simultaneous fit to two  background-subtracted 
helicity-angle distributions. The results of the fits are shown in Figures~\ref{fig:babar_rhop} 
and~\ref{fig:belle_rhop} and listed in Table~\ref{tab:summary}.

\begin{table}
\caption{\label{tab:rhop} Reconstruction requirements used in
$B^\pm \rightarrow \rho^\pm\rho^0$ analysis.
}
\renewcommand{\arraystretch}{1.1}
\begin{ruledtabular}
\begin{tabular}{lcc}
Cut & \babar\ & Belle \\
\hline
$E_\gamma$ (MeV)       &  50   &  50  (barrel); 100 (endcap) \\
$M^{}_{\gamma\gamma}$ (${\rm MeV}/c^2$) 
                       & 100. - 160.  & 118. - 150.        \\
$ M_{\pi^\pm\pi^0}$   (${\rm GeV}/c^2$)
                       &  0.396 - 1.146   & 0.65 - 0.89    \\
$ M_{\pi^+\pi^-}$   (${\rm GeV}/c^2$)
                       &  0.520 - 1.146   & 0.65 - 0.89    \\

$ \cos\theta^{}_{\!\pm}$ &   -0.9  - 0.95   &  -            \\ 
$ \cos\theta^{}_{\!0}$   &   -0.95 - 0.95   &  -            \\ 
$ p_{\pi^0}^{CM}$     (${\rm GeV}/c$)    &      &  $>0.5$    \\ 
$\mmbc$   (${\rm GeV}/c^2$)  & $5.26\!< $ & $5.272\!<$ \\
$\Delta E$   (${\rm GeV}$)   & -0.15 - 0.15  &  -0.4 - 0.4 \\
\end{tabular}
\end{ruledtabular}
\vspace{-0.4cm}
\end{table}
\begin{figure}
\includegraphics[width=0.23\textwidth]{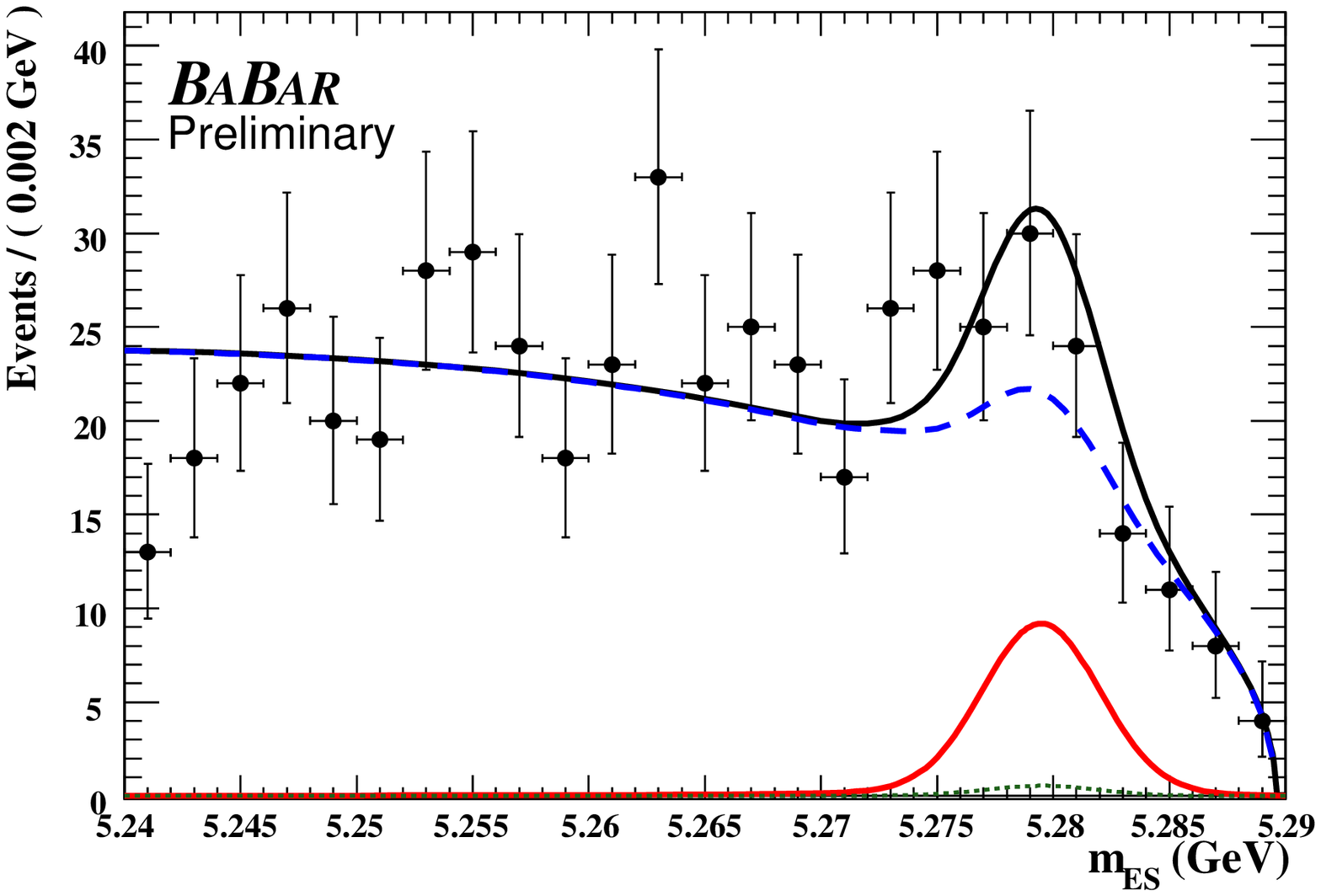}
\includegraphics[width=0.23\textwidth]{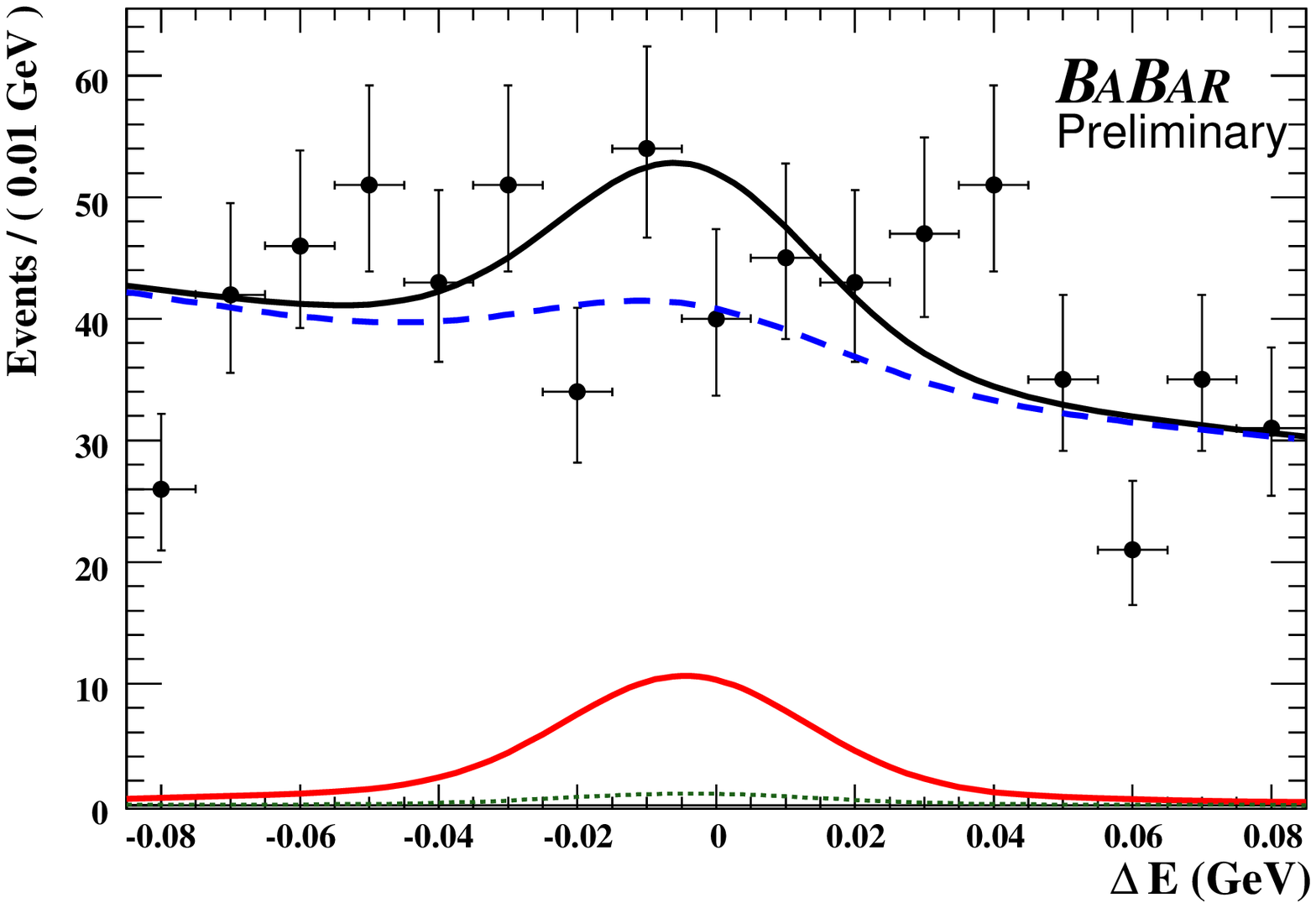}
\caption{\label{fig:babar_rho} \babar\: $m_{ES}$ and $\Delta E$ 
distributions and projections of the fit for a signal-enriched sample. 
The dashed curve is full background, the small solid curve is signal, 
the dotted curve is $B\ra \rho^0\,f_0$, and the large solid is the total.}
\vspace{-0.5cm}
\end{figure}

\begin{table*}
\caption{\label{tab:summary} Summary of $B\ra \rho\rho$ measurements.}
\renewcommand{\arraystretch}{1.4}
\begin{ruledtabular}
\begin{tabular}{l|l|ccc|ccc|c}
Decay & Quantity & \multicolumn{3}{c}{\babar} & \multicolumn{3}{c}{Belle} & HFAG \\
\cline{3-9}
           &   & Value & $N_{sig}$ & ${\cal L}$(${\rm fb}^{-1}$) &  Value & $N_{sig}$ & ${\cal L}$ 
(${\rm fb}^{-1}$)  & Value \\
\hline
   &   Br ($10^{-6}$) &   $23.5 \pm2.2\pm4.1$        &    $615\pm57$          &   316
                      &   $22.8\pm3.8^{+2.3}_{-2.6}$ &  $194\pm32$  & 253 &  $23.1^{+3.2}_{-3.3}$  \\

   &     $f_L$        &   $0.977\pm 0.024^{+0.015}_{-0.013}$   & $615\pm57$   &   316
                      &   $0.941^{+0.034}_{-0.040}\pm 0.030$   & $194\pm32$    & 253  & $0.968\pm0.23$   \\\cline{6-8}
\raisebox{2.0ex}[0pt]{$B\rightarrow \rho^\pm\rho^\mp$}   
   &     ${\cal A}_{\rho^+\rho^-}$      &  $0.07\pm 0.15\pm 0.06$  & $615\pm57$ 
      &  316
                                        &  $0.16\pm 0.21 \pm 0.07$  &  $372\pm43$     &   492  & $0.11\pm0.13$   \\
   &     ${\cal S}_{\rho^+\rho^-}$      &  $-0.19\pm 0.21^{+0.05}_{-0.07}$  & $615\pm57$  & 316          
                                        &  $0.19\pm 0.30 \pm 0.07$ &  $372\pm43$      &  492  &   $-0.06\pm0.18$   \\
\hline
   &   Br ($10^{-6}$) &   $16.8\pm2.2\pm2.3$  &  $390\pm49$ &  210.5
                      &   $31.7\pm7.1^{+3.8}_{-6.7}$ & $58.7\pm13.2$ & 78 & $18.2\pm3.0$  \\
$B\rightarrow \rho^\pm\rho^0$
   &     $f_L$        &  $0.905\pm 0.042^{+0.023}_{-0.027}$  & $390\pm49$  &   210.5  
                      &  $0.948\pm 0.106\pm0.021$  & $58.7\pm13.2$  & 78 & $0.912^{+0.044}_{-0.045}$   \\
   &     ${\cal A}_{\rho^\pm\rho^0}$      &  $-0.12\pm 0.13\pm 0.10$  & $390\pm49$  & 210.5    
                                          & $0.00\pm 0.22 \pm 0.03$   & $58.7\pm13.2$  & 78 & $-0.08\pm0.13$   \\
\hline
   &   Br ($10^{-6}$) &   $1.16^{+0.37}_{-0.36}\pm0.27$  & $98^{+32}_{-31}$ & 316   
                      &                  &     &     & $1.16\pm0.46$  \\
\raisebox{2.0ex}[0pt]{$B\rightarrow \rho^0\rho^0$}   &     $f_L$   &  $0.86^{+0.11}_{-0.13}\pm0.05$ & 
                                            $98^{+32}_{-31}$ & 316  &    &   &  &  $0.86^{+0.12}_{-0.14}$  \\
\end{tabular}
\end{ruledtabular}
\vspace{-0.4cm}
\end{table*}
\vspace{-0.4cm}

\section{$B^0 \rightarrow \rho^0\rho^0$}
\babar\ finds evidence for $B^0 \rightarrow \rho^0\rho^0$ decays and measures 
its branching fraction and polarization using a sample of about 348 million 
$B\overline B$ pairs~\cite{babar_rho0rho0}. Events are selected from the 
region $5.24~{\rm GeV}/c^2\!<\!\mmbc\!<\!5.29~{\rm GeV}/c^2$, 
$|\Delta E|\!<\!85~{\rm MeV}$, and are required to satisfy 
$0.55 <\! M_{\pi^+\pi^-}\!<\!1.05~{\rm GeV}/c^2$, and $|\cos\theta_0|<\!0.98$.
In events with multiple $B$ candidates one is selected based on the the best $\chi^2$ of 
a four-pion vertex fit. Additional suppression of the dominant continuum background is
achieved using the flavor tagging information. The data sample is divided into seven 
tag-quality intervals, $c_{tag}$. The $B^0\rightarrow \rho^0\rho^0$ event yield and 
polarization $f_L$ are obtained from an unbinned extended ML fit to $m_{ES}$, 
$\Delta E$, ${m_{\pi^+\pi^-}}_{1,2}$, ${\cos\theta}_{1,2}$, ${\cal N}$, 
and $c_{tag}$ distribution of 65180 events. The fit also allows to obtain the yields
for $B^0\ra \rho^0\,f_0(990)$ and $B^0\ra f_0(980)\,f_0(980)$ decays.
The charmless background is  dominated by  $B^0\ra  a_1^\pm\pi^\mp$ events which number is 
a free parameter in the fit. Other $b\rightarrow u$ background modes include: 
$B\ra (\rho^0\,K^{*0}, \rho^+\,\rho^0,\rho\,\pi)$, and $B^0\ra\rho^+\,\rho^-$. 
The fit results are shown in Fig.~\ref{fig:babar_rho}  and listed in  Table~\ref{tab:summary}.

\section{Constraint on {\boldmath $\alpha (\phi_2$)}}
We constrain $\phi^{}_2$ using an isospin analysis~\cite{gronau_london}, which 
allows one to relate six observables to six underlying parameters: five decay 
amplitudes for $B\ra\rho\rho$  and the angle $\phi_2$.
The observables are the branching fractions for $B\ra\rho^+\rho^-$, $\rho^+\rho^0$, 
and $\rho^0\rho^0$ (listed in Table~\ref{tab:summary})\,\cite{hfag}; the $CP$ parameters 
${\cal A}_{\rho^+\rho^-}$ and ${\cal S}_{\rho^+\rho^-}$; and the parameter ${\cal A}_{\rho^0\rho^0}$ 
for $B\ra\rho^0\rho^0$ decays. The branching fractions must be multiplied by the 
corresponding longitudinal polarization fractions (taken from Table~\ref{tab:summary})~\cite{hfag}. 
We neglect possible contributions from electroweak penguins and  $I\!=\!1$ amplitudes~\cite{falk} 
to \brhorho. We follow the statistical method of Ref.~\cite{charles} and construct a  $\chi^2({\phi_2})$ 
using the measured values and  obtain a minimum $\chi^2$ (denoted $\chi^2_{\rm min}$); 
we then scan \phitwo\ from 0$^\circ$ to 180$^\circ$, calculating the difference 
$\Delta\chi^2\equiv\chi^2(\phi^{}_2)-\chi^2_{\rm min}$. We insert $\Delta\chi^2$ into 
the cumulative distribution function for the $\chi^2$ distribution for one degree of 
freedom to obtain a confidence level (C.L.) for each \phitwo\ value. 
\begin{figure}
\includegraphics[width=0.4\textwidth]{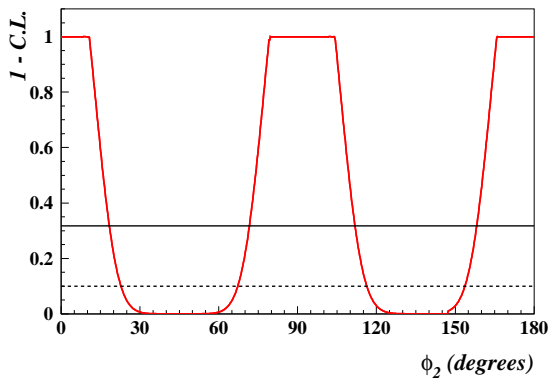}
\caption{\label{fig:phi2} 1 - C.L. vs. $\phi_2$. The horizontal lines denote C.L. = 
$68.3\%$ (solid) and C.L. = $90\%$ (dashed)}
\vspace{-0.3cm}
\end{figure}
The resulting function $1\!-\!{\rm C.L.}$ (Fig.~\ref{fig:phi2}) has more than one peak 
due to ambiguities that arise when solving for $\phi^{}_2$. Because ${\cal A}_{\rho^0\rho^0}$ 
is not yet measured, we allow this observable to float; this produces the ``flat-top'' 
regions in Fig.~\ref{fig:phi2}. The solution consistent with the Standard Model is 
$71^\circ\!<\!\phi^{}_2\!<\!113^\circ$ at 68\%~C.L. or 
$67^\circ\!<\!\phi^{}_2\!<\!116^\circ$ at 90\%~C.L. Recently, a different model-dependent 
approach to extract $\phi_2$ using flavor $SU$(3) symmetry has been proposed~\cite{alpha_su3}. 
This method would give more stringent constraints on $\phi_2$.

In summary, we present recent measurements in $B\ra \rho\rho$ decays.
These measurements are used to constrain the angle~$\phi^{}_2$ using an 
isospin analysis.

\end{document}